\documentclass[letterpaper]{IEEEtran}

\usepackage[ colorlinks = true,
             linkcolor = blue,
             urlcolor  = blue,
             citecolor = red,
             anchorcolor = magenta,
]{hyperref}

\usepackage{amssymb,amsmath,amsthm}
\usepackage{cite}

\theoremstyle{plain}  
\newtheorem{lemma}{Lemma}

\newtheorem{proposition}[lemma]{Proposition}
\newtheorem{corollary}[lemma]{Corollary}
\newtheorem{theorem}[lemma]{Theorem}

\theoremstyle{definition}

\newcommand{\cX}{\mathcal{X}}

\newcommand{\cP}{\mathcal{P}}
\newcommand{\cS}{\mathcal{S}}
\newcommand{\cC}{\mathcal{C}}

\newcommand{\cB}{\mathcal{B}}
\newcommand{\cE}{\mathcal{E}}

\newcommand{\cD}{\mathcal{D}}
\newcommand{\cN}{\mathcal{N}}
\newcommand{\cH}{\mathcal{H}}

\newcommand{\proj}[1]{|k\rangle\!\langle k|}

\DeclareMathOperator{\tr}{tr}

\newcommand{\RainsSet}{{\rm PPT}'}
\newcommand{\Sepp}{{\rm SEP}}
\newcommand{\PPT}{{\rm PPT}}
\newcommand{\CPP}{{CPPP }}

\newcommand{\CPPsub}{\rm pp}

\begin{document}
\flushbottom

\title{Strong Converse Rates for Quantum Communication}
 
 \author{Marco Tomamichel$^*$, \and Mark M. Wilde$^\dagger$, \and Andreas Winter$^\ddagger$%
 \thanks{$^*$ School of Physics, The University of Sydney, Sydney, NSW 2006, Australia. (Email: {marco.tomamichel@sydney.edu.sg}).}%
 \thanks{$^\dagger$ Hearne Institute for Theoretical Physics, Department of
Physics and Astronomy, Center for Computation and Technology, Louisiana State
University, Baton Rouge, Louisiana 70803, USA.}%
\thanks{$^\ddagger$ ICREA \& F\'isica Te\'orica, Informaci\'o i Fenomens
Qu\'antics, Universitat Aut\`{o}noma de Barcelona, ES-08193 Bellaterra
(Barcelona), Spain.}}
   

\maketitle

\begin{abstract}
We revisit a fundamental open problem in quantum information theory, namely whether it is possible to transmit quantum information at a rate exceeding the channel capacity if we allow for a non-vanishing probability of decoding error. 
Here we establish that the Rains information of any quantum channel is a strong
converse rate for quantum communication: For any sequence of codes with rate exceeding the Rains information of the channel, we show that the fidelity vanishes exponentially fast as the number of channel uses increases. This remains true even if we consider codes that perform classical post-processing on the transmitted quantum data.
As an application of this result, for generalized dephasing channels we show that the Rains information is also achievable, and thereby establish the strong converse property for quantum communication over such channels. Thus we conclusively settle the strong converse question for a class of quantum channels that have a non-trivial quantum capacity.
\end{abstract}

\section{Introduction}
 
The \emph{quantum capacity} of a quantum channel $\cN$, denoted $Q(\mathcal{N})$, is defined as the maximum rate (in qubits per channel use) at
which it is possible to transmit quantum information over many memoryless uses of the channel with a fidelity that asymptotically converges to one as we increase the number of channel uses. (We will formally introduce capacity in Section~\ref{sec:ent-gen-code}.)
The question of determining
the quantum capacity was set out by Shor in his seminal
paper on quantum error correction \cite{S95}. Since then, a number of works
established a \textquotedblleft multi-letter\textquotedblright\ upper bound on
the quantum channel capacity in terms of the coherent information
\cite{SN96,BKN98,BNS98}, and the coherent information lower bound on quantum
capacity was demonstrated by a sequence of works
\cite{L97,capacity2002shor,D05} which are often said to bear \textquotedblleft
increasing standards of rigor.\textquotedblright\footnote{However, see the
later works \cite{qcap2008second} and \cite{qcap2008fourth}, which set
\cite{L97} and \cite{capacity2002shor}, respectively, on a firm foundation.}
In more detail, the work in \cite{L97,capacity2002shor,D05}\ showed that the
following \emph{coherent information} of the channel$~\mathcal{N}$ is an \emph{achievable rate
for quantum communication}:%
\begin{equation}
I_{\rm c}(  \mathcal{N})  :=  \sup_{\phi_{RA}} I(R\rangle B)_{\rho},  \quad \textrm{where} \quad \rho_{RB} = \mathcal{N}_{A\rightarrow B}(  \phi_{RA})
\label{eq:coh-info}%
\end{equation}
and the optimization is over all pure bipartite states $\phi_{RA}$. Here, the
coherent information of a bipartite state $\rho_{RB}$ is defined as
$I(R\rangle B)_{\rho}:= H(\rho_B)-H(\rho_{RB})$, with the von Neumann entropies $H(\rho) :=
-$Tr$\left\{  \rho \log\rho \right\}$.\footnote{All logarithms in this paper are taken base two.} From the above
result, we can also conclude that the rate $I_{\rm c}(  \mathcal{N}^{\otimes
\ell})  /\ell$ is achievable for any positive integer~$\ell$, simply by applying
the formula in (\ref{eq:coh-info}) to the \textquotedblleft
superchannel\textquotedblright\ $\mathcal{N}^{\otimes \ell}$ and normalizing. By
a limiting argument, we find that the regularized coherent information
$\lim_{\ell\rightarrow\infty}I_{c}(  \mathcal{N}^{\otimes \ell})  /\ell$ is
also achievable, and Refs.~\cite{SN96,BKN98,BNS98} established that this
regularized coherent information is also an upper bound on quantum capacity. This establishes that
\begin{align}
  Q(\mathcal{N}) = \lim_{\ell\rightarrow\infty} \frac{I_{\rm c}(  \mathcal{N}^{\otimes \ell})}{\ell} . \label{eq:capaity-result}
\end{align}

Clearly, the regularized coherent information is not a tractable
characterization of quantum capacity. But the later work of Devetak and Shor
proved that the \textquotedblleft single-letter\textquotedblright\ coherent
information formula in (\ref{eq:coh-info}) is equal to the quantum capacity
for the class of degradable quantum channels \cite{DS05}. Degradable channels
are such that the receiver of the channel can simulate the channel to the
environment by applying a degrading map to the channel output.

All of the above works established an understanding of quantum capacity in the
following sense:

\begin{enumerate}
\item (Achievability) If the rate of quantum communication is below the
quantum capacity, 
then there exists a scheme for quantum communication such
that the fidelity converges to one in the limit of many channel uses.

\item (Weak Converse) If the rate of quantum communication is above the
quantum capacity, 
then there cannot exist an asymptotically error-free quantum
communication scheme.
\end{enumerate}

It is crucial to note at this point that practical schemes for quantum communication are restricted to operate on finite block lengths, and it is thus in principle impossible to achieve exactly error-free communication (for most channels). Hence, 
we are left to wonder whether it is possible to transmit information at a rate larger  than the regularized coherent information given in~\eqref{eq:capaity-result} if a non-vanishing error is permissible. 
Interestingly, it has been known
since the early days of classical information theory that the
capacity of a classical channel obeys the \emph{strong converse property}
\cite{Wolfowitz1964,Arimoto73}: if the rate of communication exceeds the capacity,
then the error probability necessarily converges to one in the limit of many channel uses. 
Furthermore, many works have now confirmed that the strong
converse property holds for the classical capacity of several quantum channels
\cite{ON1999,W99,KW09,WWY13,WW13} and also for the entanglement-assisted
classical capacity of all quantum channels \cite{BDHSW12,BCR09,GW13}.

Consequently, the present work aims
to sharpen the interpretation of quantum capacity in the same spirit.

\subsection{Overview of Results and Outline}

In this paper, we extend the Rains relative entropy of a quantum state \cite{R01,AdMVW02} by defining the \emph{Rains information of a quantum channel}
$\mathcal{N}_{A\rightarrow B}$\ as follows:\footnote{It should be clear from
the context whether $R$ refers to ``Rains'' or to a reference system.}%
\begin{equation}
R(  \mathcal{N})  := \sup_{\rho_{RA}} \inf_{\tau_{RB}\in\RainsSet(
R:B)  }D(  \cN_{A \to B}(\rho_{RA}) \Vert\tau_{RB})  , \label{eq:rains-bound}
\end{equation}
where we maximize over input states $\rho_{RA}$ with $R$ a reference system, and minimize over a set of subnormalized states (including all PPT states) first considered by Rains, namely the set
\begin{align}
\RainsSet(  R\!:\!B)   &  =\left\{  \tau_{RB}:\tau_{RB}%
\geq0\wedge\left\Vert T_{B}(  \tau_{RB})  \right\Vert _{1}%
\leq1\right\} .%
\end{align}
Here, $D(\rho\|\tau) := \tr \{ \rho (\log \rho - \log \tau) \}$ is the relative entropy and $T_{B}$ denotes the partial transpose. The quantity in (\ref{eq:rains-bound}), often
called the \textquotedblleft Rains bound\textquotedblright, was first
explored by Rains in the context of entanglement distillation \cite{R01} and
later refined to the form above in \cite{AdMVW02}. 

Our main contribution is that the Rains information of a quantum channel is a \emph{strong converse rate for quantum communication}
(Theorem~\ref{thm:strong-converse}), even if we allow for classical pre- and post-processing. (The allowed codes are described in more detail in Section~\ref{sec:ent-gen-code}.) That is, if the quantum
communication rate of any protocol for a channel $\mathcal{N}_{A\rightarrow
B}$ exceeds its Rains information, then the fidelity of the scheme decays
exponentially fast to zero as the number $n$ of channel uses increases. This result thus provides an operational proof of the inequality $I_{\rm c}(\mathcal{N}) \leq R(\mathcal{N})$. Note however that Theorem~4 of \cite{PVP00} provides a direct proof of this inequality
via operator monotonicity of the logarithm and the notion of the reduction criterion from \cite{PhysRevA.59.4206}.  

We
establish this strong converse theorem by exploiting the generalized divergence framework of
Sharma and Warsi \cite{SW12}, which generalizes the related framework from the classical context
\cite{PV10} (see also earlier developments from
\cite{N01}).
However, as discussed in Section~\ref{sec:gen-div}, our main departing point from that work is to consider
a different class of \textquotedblleft useless\textquotedblright\ channels for
quantum data transmission. That is, in \cite{SW12}, the main idea was to
exploit a generalized divergence to compare the output of the channel with an
operator of the form $\pi_{R}\otimes\sigma_{B}$, where $\pi_R$ denotes the fully mixed state on the reference system $R$ and $\sigma_B$ is an arbitrary state on the output system. This state can be viewed as the output of a
\textquotedblleft useless\textquotedblright\ channel that replaces the input
and reference system with the maximally mixed state and an uncorrelated state. The resulting
information quantity is a generalization of the coherent information from
(\ref{eq:coh-info}), and one then invokes a data processing inequality to
relate this quantity to communication rate and fidelity. Here, we instead
compare the output of the channel with an operator in the set $\RainsSet%
(  R\!:\!B)  $, which contains and is closely related to the positive
partial transpose states, which in turn are well known to have no distillable
entanglement (or equivalently, no quantum data transmission capabilities, so
that they constitute another class of \textquotedblleft
useless\textquotedblright\ channels for quantum data transmission). 

It turns
out that a R\'{e}nyi-like version of the Rains information of a quantum
channel appears to be easier to manipulate, so that we can show that it obeys
a weak subadditivity property (Theorem~\ref{thm:weak-subadditivity} in Section~\ref{sec:weaksub}). From
there, some standard limiting arguments conclude the proof of
our main result (Theorem~\ref{thm:strong-converse} in Section~\ref{sec:main}).

The main application of this result is to establish the \emph{strong converse property} for any generalized dephasing channel (Section~\ref{sec:dephasing}). The
action of any channel in this class on an input state $\rho$ is as follows:%
\begin{equation}
\mathcal{N}(  \rho)  =\sum_{x,y=0}^{d-1}\left\langle
x\right\vert _{A}\rho\left\vert y\right\rangle _{A}\ \left\langle \psi
_{y}|\psi_{x}\right\rangle \ \vert x\rangle \langle
y\vert _{B},
\end{equation}
where $\{\vert x\rangle _{A}\}$ and $\{\vert x\rangle
_{B}\}$ are orthonormal bases for the input and output systems, respectively,
for some positive integer $d$, and $\{\left\vert \psi_{x}\right\rangle _{E}\}$
is a set of arbitrary pure quantum states. A particular example in this class
is the qubit dephasing channel, whose action on a qubit density operator is%
\begin{equation}
\mathcal{N}(  \rho)  =\left(  1-p\right)  \rho+pZ\rho Z,
\end{equation}
where the dephasing parameter $p\in\left[  0,1\right]  $ and $Z$ is the Pauli
$\sigma_Z$ operator. We prove this result by showing that the Rains information of a
generalized dephasing channel is equal to its coherent information
(Proposition~\ref{prop:I_c=R-for-H}). This extends \cite[Theorem~6.2]{R01} from ``maximally correlated states'' to generalized dephasing channels.

Finally, in Section~\ref{sec:erasure}, we discuss how our technique implies
the strong converse property for the classical post-processing assisted
quantum capacity of the quantum erasure channel. Note however that this result
was previously established in \cite{BBCW13}.

\subsection{Related Work}

A few papers have made some partial progress on or addressed the strong
converse for quantum capacity question
\cite{BDHSW12,BCR09,BBCW13,MW13,SW12,WW14}, with none however establishing
that the strong converse holds for any nontrivial class of channels.
Refs.~\cite{BDHSW12,BCR09} prove a strong converse theorem for the
entanglement-assisted quantum capacity of a channel, which in turn establishes
this as a strong converse rate for the unassisted quantum capacity.
Ref.~\cite{BBCW13} proves that the entanglement cost of a quantum channel is a
strong converse rate for the quantum capacity assisted by unlimited forward
and backward classical communication, which then demonstrates that this
quantity is a strong converse rate for unassisted quantum capacity. 

Arguably the most
important progress to date for establishing a strong converse for quantum
capacity is from \cite{MW13}. These authors reduced the proof of the strong
converse for the quantum capacity of degradable channels to that of
establishing it for the simpler class of channels known as the symmetric
channels (these are channels symmetric under the exchange of the receiver and
the environment of the channel). Along the way, they also demonstrated that a
\textquotedblleft pretty strong converse\textquotedblright\ holds for
degradable quantum channels, meaning that there is (at least) a jump in the quantum error
from zero to $1/2$ as soon as the communication rate exceeds the quantum
capacity. 

We also mention that our main contribution here sets a previous claim from \cite{SSW08} on a firm
foundation. Namely, in the introduction of Ref.~\cite{SSW08}, the authors
claim that the Rains bound for entanglement distillation leads to a weak
converse upper bound on the quantum capacity of any channel. However,
Ref.~\cite{SSW08} does not appear to sufficiently support this claim. In particular, the authors do not provide an argument for the weak sub-additivity of the Rains information of a quantum
channel. (We establish the corresponding property for R\'enyi Rains information of a quantum channel in Theorem~\ref{thm:weak-subadditivity} in Section~\ref{sec:weaksub}.)

Sharma and Warsi established the \textquotedblleft
generalized divergence\textquotedblright\ framework for understanding quantum
communication and reduced the task of establishing the strong converse to a
purely mathematical \textquotedblleft additivity\textquotedblright\ question
\cite{SW12}, which hitherto has remained unsolved. The later work in
\cite{WW14}\ demonstrated that randomly
selected codes with a communication rate exceeding the quantum capacity of the
quantum erasure channel lead to a fidelity that decreases exponentially fast as the number of channel uses increases.
(We stress that a strong converse would imply this behavior  for \textit{all} codes whose rate exceeds capacity.)

The methods given in the present paper also imply that the Rains relative entropy
of a quantum state is a strong converse rate for entanglement distillation, as established by Hayashi in \cite[Section~8.6]{H06} using a different method.


\section{Preliminaries}
\label{sec:notation}

\subsection{States and Channels} 

\paragraph*{States}
We denote different physical systems by capital letters (e.g.\ $A$, $B$) and we use these labels as subscripts to indicate  with which physical system a mathematical object is associated.
Let $\cH_A$ denote the Hilbert space corresponding to the system $A$, where we restrict ourselves to
finite-dimensional Hilbert spaces throughout this paper. We define $|A| = \dim \{ \cH_A \}$. Moreover, let $\mathcal{B}(A)$ and $\mathcal{P}(A)$ denote the algebra of \emph{bounded linear
operators} acting on $\mathcal{H}_A$ and the subset of \emph{positive semi-definite operators}, respectively.
We also write
$X_A \geq0$ if $X_A \in\mathcal{P}(A)$. 
An\ operator $\rho_A$
is in the set $\mathcal{S}( A)  $\ of \emph{quantum states}
if $\rho_A \geq 0$ and $\tr \left\{
\rho_A \right\} =1$. We say that a quantum state $\rho_A$ is \emph{pure} if $\text{rank}(\rho_A) = 1$. We denote the \emph{identity operator} in $\cB(A)$ by $I_A$ and the \emph{fully mixed} state by $\pi_A = I_A / |A|$.

The fidelity between two density operators $\rho$ and $\sigma$
is defined as \cite{U76}%
\begin{equation}
F(  \rho,\sigma)  :=\left\Vert \sqrt{\rho}\sqrt{\sigma
}\right\Vert _{1}^{2} = \left[\tr\Big\{ \sqrt{\sqrt{\rho}\,\sigma\sqrt{\rho}\,} \Big\}\right]^2.
\end{equation}

A bipartite physical system $AB$ is described by the tensor-product Hilbert space $\cH_{AB} = \mathcal{H}_{A}\otimes\mathcal{H}_{B}$ with the sets $\mathcal{B}(AB)$, $\cP(AB)$, and $\cS(AB)$ defined accordingly. Given a
bipartite quantum state $\rho_{AB} \in\cS(AB)$, we unambiguously write $\rho_{A}=\tr_{B}\left\{
\rho_{AB}\right\}$ for the reduced state on system $A$. A state $\rho_{AB}$ is called \emph{maximally entangled}
with Schmidt rank $M$ if there exist orthonormal bases $\{\vert x \rangle_A\}$ and
$\{\vert x \rangle_B\}$ such that
$\rho_{AB} = \vert \Phi \rangle \langle \Phi \vert_{AB}$, where
\begin{equation}
\vert \Phi \rangle_{AB} = \frac{1}{\sqrt{M}}
\sum_{x = 1}^{M} \vert x \rangle_A \otimes \vert x \rangle_B.
\end{equation}
A product state is a state that can be written in the form $\rho_{AB} = \rho_A \otimes \rho_B$.

A state $\rho_{AB}$ is called \emph{separable} if it can be written in the form
\begin{align}
  \rho_{AB} = \sum_{x \in \cX} P(x) \rho_A^x \otimes \rho_B^x \,
\end{align}
for a set $\cX$, a probability mass function $P$ on $\cX$ and $\rho_A^x \in \cS(A)$, $\rho_B^x \in \cS(B)$ for all $x \in \cX$ \cite{W89}. We denote the set of all such states by $\Sepp(A\!:\!B)$.
Furthermore, we say that a bipartite state $\rho_{AB}$\ is PPT if it has a positive partial
transpose, namely if $T_{B}(  \rho_{AB})  \geq0$, where $T_{B}$ indicates the partial transpose operation on system $B$. This is a necessary condition for a bipartite state to be
separable~\cite{P96}.\footnote{It is also sufficient for $2\times2$ and $2\times3$ systems, but
otherwise only necessary \cite{Horodecki19961}.} Let $\PPT(  A\!:\!B)  $
denote the set of all such states.\footnote{Note that there is no need to have an asymmetric notation here because a state is PPT\ with respect to transpose on
system $A$ if and only if it is PPT\ with respect to a partial transpose on
system $B$.}

For our purposes it is useful to further enlarge $\PPT(A\!:\!B)$ to the set $\RainsSet(A\!:\!B)$ \cite{AdMVW02}, defined as
\begin{equation}
\RainsSet(  A:B)  :=\left\{  \tau_{AB}:\tau_{AB}\geq
0\wedge\left\Vert T_{B}(  \tau_{AB})  \right\Vert _{1}%
\leq1\right\}  .
\end{equation}
The set $\RainsSet(  A\!:\!B)$\ includes all PPT\ states because
$\left\Vert T_{B}(  \tau_{AB})  \right\Vert _{1}=1$ if $\tau
_{AB}\in\ \PPT(  A\!:\!B)  $. All operators $\tau_{AB}%
\in\RainsSet(  A\!:\!B)  $ are subnormalized, in the sense that
Tr$\left\{  \tau_{AB}\right\}  \leq1$, because%
\begin{equation}
\tr\left\{  \tau_{AB}\right\}  =\tr\left\{  T_{B}(  \tau
_{AB})  \right\}  \leq\left\Vert T_{B}(  \tau_{AB})
\right\Vert _{1}\leq1.
\end{equation}
However, the set $\RainsSet$ contains strictly sub-normalized states that are not PPT.

\paragraph*{Channels}

We denote linear maps from the set of bounded linear operators on one system to the bounded linear operators on another system by calligraphic letters. For example, $\cN_{A\to B}$ denotes a map from $\cB(A)$ to $\cB(B)$, and we will drop the subscript $A \to B$ if it is clear from the context.
Let id$_{A}$ denote the \emph{identity map} acting on $\cB(A)$. 
If a linear map is completely positive and
trace-preserving (CPTP), we say that it is a \emph{quantum channel}. 

We say that a CPTP map $\cN_{AB \to A'B'}$ consists of local operations (LO) from $A\!:\!B$ to $A'\!:\!B'$ if it has the form $\cN_{AB \to A'B'} = \mathcal{L}_{A \to A'} \otimes \mathcal{M}_{B \to B'}$ for CPTP maps
$\mathcal{L}_{A \to A'}$ and $\mathcal{M}_{B \to B'}$.
Similarly, we say that $\cN$ is LOCC if it consists of local operations and classical communication \cite{BDSW96,CLMOW14}. Furthermore, LOCC maps are contained in the set of separability preserving maps from $A\!:\!B$ to $A'\!:\!B'$ that take $\Sepp(A\!:\!B)$ to $\Sepp(A'\!:\!B')$.
Finally, $\mathcal{N}$ is a
PPT preserving operation from $A\!:\!B$ to $A'\!:\!B'$ if the map $T_{B^{\prime}}\circ\mathcal{N}%
_{AB\rightarrow A^{\prime}B^{\prime}}\circ T_{B}$ is CPTP~\cite{R01}. (So we can conclude that any LOCC map is a PPT-preserving map.)

\subsection{Codes, Rates and Capacity}
\label{sec:ent-gen-code}

\paragraph*{Codes}
We consider a general class of classical pre- and post-processing (CPPP) assisted {entanglement generation (EG)
codes}, for which the goal is for the sender (Alice) to use a quantum channel in order to
share a state that is close to a maximally entangled state with the receiver (Bob).\footnote{This generalizes the standard classical scenario in which we are interested in transmitting a uniformly distributed message.} The \CPP assisted codes allow classical pre- and post-processing before and after the quantum communication phase. In particular, the parties are allowed to prepare a separable resource state before the quantum communication commences.
Note that any converse bounds for these classical communication assisted codes naturally also imply the same bounds for unassisted codes.

Formally, we define a \emph{\CPP assisted EG code} for a channel~$\mathcal{N}_{A \to B}$ as a triple 
\begin{align}
\cC = (M, \cE_{A_0 B_0 \to \tilde{A}A\tilde{B}}, \cD_{\tilde{A}B\tilde{B} \to \hat{A}\hat{B}}) \,. \label{eq:code}
\end{align}
Here, $A_0 \cong B_0 \cong \mathbb{C}$ are trivial, $\hat{A}$ and $\hat{B}$ are Hilbert spaces of dimension $M$, and 
$\tilde{A}$ and $\tilde{B}$ are auxiliary Hilbert spaces of arbitrary dimension. Moreover, $\cE_{A_0B_0 \to \tilde{A}A\tilde{B}}$ is an LOCC quantum channel from $A_0\!:\!B_0$ to $\tilde{A}A\!:\!\tilde{B}$, and $\cD_{\tilde{A}\tilde{B}B \to \hat{A}\hat{B}}$ is an LOCC quantum channel from $\tilde{A}\!:\!\tilde{B}B$ to $\hat{A}\!:\!\hat{B}$. We write $|\cC| = M$ for the \emph{size} of the EG code. An \emph{unassisted EG code} is defined in the same way, but $\cE$ and $\cD$ are restricted to be LO instead of LOCC.

The corresponding
coding schemes thus begin with Alice and Bob preparing a bipartite state $\rho_{\tilde{A}A\tilde{B}} \in \Sepp(\tilde{A}A\!:\!\tilde{B})$ using the quantum channel $\cE$. (This state is restricted to be a product state for unassisted codes but can be an arbitrary separable state in the CPPP assisted case.)
Alice then sends the system $A$ through the channel $\mathcal{N}$, resulting in the state
\begin{equation}
\rho_{\tilde{A}B\tilde{B}} = \mathcal{N}_{A\rightarrow B}(  \rho_{\tilde{A}A\tilde{B}})  .
\label{eq:state-output-from-channel}%
\end{equation}
Finally, Alice and Bob perform a decoding $\mathcal{D}$, leading
to the state
$\omega_{\hat{A}\hat{B}} = \mathcal{D}_{\tilde{A}B\tilde{B}\rightarrow\hat{A}\hat{B}}(  \rho_{\tilde{A}B\tilde{B}}  )$.

The \emph{fidelity} of the above code $\cC$ on the channel $\cN$ is given by%
\begin{equation}
F(\cC, \cN) := \left\langle \Phi\right\vert _{\hat{A}\hat{B}}\, \omega_{\hat{A}\hat{B}}\, \left\vert
\Phi\right\rangle _{\hat{A}\hat{B}}, \label{eq:fidelity}%
\end{equation}
where $\left\vert \Phi\right\rangle _{\hat{A}\hat{B}}$ is a (fixed) maximally entangled
state on $\hat{A}\hat{B}$ of Schmidt rank $M$.

\paragraph*{Rates And Capacity}
The main focus of this paper is on EG codes for many parallel uses of a
memoryless quantum channel. That is, we want to investigate product channels $\mathcal{N}^{\otimes n}$ for large $n$. Note that a \CPP assisted EG code for $\cN^{\otimes n}$ (as described above) allows for classical communication before and after the product channel is used, but does not allow for interactive schemes with classical communication between different channel uses.

A rate $r$ is an \emph{achievable rate} for (\CPP assisted) quantum communication over the channel $\cN$ if there exists a sequence of codes $\{\cC_n \}_{n \in \mathbb{N}}$ where $\cC_n$ is a (\CPP assisted) EG code for $\cN^{\otimes n}$, such that
\begin{align}
  \liminf_{n\to\infty} \frac{1}{n} \log |\cC_n| & \geq r ,\\
   \lim_{n\to\infty} F(\cC_n, \cN^{\otimes n}) & = 1 .
\end{align}
The \emph{quantum capacity} of $\cN$, denoted $Q(\cN)$  is the supremum of all  achievable rates. Analogously, the \emph{\CPP assisted quantum capacity} of $\cN$, denoted $Q_{\CPPsub}(\cN)$, is the supremum of all \CPP assisted achievable rates.

On the other hand, $r$ is a \emph{strong converse rate} for (\CPP assisted) quantum communication if for every sequence of codes $\{ C_n \}_{n \in \mathbb{N}}$ 
as above, we have
\begin{align}
  \liminf_{n\to\infty} \frac{1}{n} \log |\cC_n| > r \, \implies \,\lim_{n\to\infty} F(\cC_n, \cN^{\otimes n}) = 0 .
\end{align}
The \emph{strong converse quantum capacity}, denoted $Q^{\dagger}(\cN)$, is the infimum of all strong converse rates. Analogously, the \emph{\CPP assisted strong converse quantum capacity} of $\cN$, denoted $Q_{\CPPsub}^{\dagger}(\cN)$, is the infimum of all \CPP assisted strong converse rates. 

Clearly, the following inequalities hold by definition:
\begin{align}
Q(\cN) & \leq Q_{\CPPsub}(\cN) \leq Q_{\CPPsub}^{\dagger}(\cN) ,\\
 Q(\cN) & \leq  Q^{\dagger}(\cN) \leq Q_{\CPPsub}^{\dagger}(\cN) \,.
\end{align}
Finally, we say that a channel $\cN$ satisfies the \emph{strong converse property for quantum communication} if $Q(\cN) = Q^{\dagger}(\cN)$. Similarly, we say that a channel $\cN$ satisfies the \emph{strong converse property for \CPP assisted quantum communication} if $Q_{\CPPsub}(\cN) = Q_{\CPPsub}^{\dagger}(\cN)$.


\section{Generalized Divergence Framework}
\label{sec:gen-div}

A functional $\mathbf{D}:\mathcal{S}  \times \mathcal{P}
\rightarrow \mathbb{R}$ is a generalized divergence if it satisfies the monotonicity inequality%
\begin{equation}
\mathbf{D}(  \rho \Vert\sigma)  \geq\mathbf{D}(  \mathcal{N}%
(  \rho)  \Vert\mathcal{N}(  \sigma)  )  ,
\end{equation}
where $\mathcal{N}$ is a CPTP map. It follows
directly from monotonicity that any generalized divergence is invariant under
isometries, in the sense that
$\mathbf{D}(  \rho\Vert\sigma)  =\mathbf{D}(U\rho U^{\dag
}\Vert U\sigma U^{\dag})$, where $U$ is an isometry, and that it is invariant under tensoring with
another quantum state $\tau$, namely $\mathbf{D}(  \rho\Vert\sigma)  =\mathbf{D}(\rho \otimes\tau\Vert\sigma\otimes\tau)$. Note that to establish isometric invariance from monotonicity, we require a channel that can reverse the action of an isometry (see, e.g., 
\cite[Section~4.6.3]{W15book} for this standard construction).

\subsection{Rains Relative Entropy and Rains Information}

We now define some information measures which play a central role in this
paper. They are direct generalizations of the Rains bound on distillable entanglement
\cite{R01} and the subsequent reformulation of it in \cite{AdMVW02}. 

We define
the \emph{generalized Rains relative entropy} of a bipartite state $\rho_{AB}$ as
follows:%
\begin{equation}
R_{\mathbf{D}}(A\!:\!B)_{\rho}  :=\inf_{\tau_{AB}\in
\RainsSet(  A:B)  }\mathbf{D}(  \rho_{AB}\Vert\tau
_{AB})  ,
\end{equation}
We sometimes abuse notation and write $R_{\mathbf{D}}(A\!:\!B)_{\rho} = R_{\mathbf{D}}(\rho_{AB})$ if the bipartition is obvious in context.

One property of $R_{\mathbf{D}}$, critical for our
application here, is that it is monotone under PPT preserving operations, in
the sense that
\begin{equation}
R_{\mathbf{D}}(A\!:\!B)_{\rho}  \geq R_{\mathbf{D}}(A'\!:\!B')_{\omega} ,  
\end{equation}
where $\omega_{A'B'} = \mathcal{P}_{AB\to A'B'}(  \rho_{AB})$
and $\mathcal{P}_{AB\to A'B'}$ is any PPT-preserving operation from $A\!:\!B$ to $A'\!:\!B'$. This is because
PPT-preserving operations do not take operators $\tau_{AB}$ in $\RainsSet%
(  A\!:\!B)  $ outside of this set, which follows from
\begin{align}
\Vert T_{B'} (\mathcal{P}_{AB\to A'B'}(\tau_{AB})) \Vert_{1} & = \Vert T_{B'}
(\mathcal{P}_{AB}(T_{B} (T_{B}(\tau_{AB})))) \Vert_{1}
\nonumber\\
& \leq\Vert T_{B}%
(\tau_{AB}) \Vert_{1} \leq1.
\end{align}
In the above, the equality follows because the partial transpose $T_{B}$
is its own inverse, and the first inequality follows because the map $T_{B'}
\circ\mathcal{P}_{AB} \circ T_{B}$ is CPTP (as $\mathcal{P}_{AB}$ is PPT preserving) and the
fact that the trace norm is monotone decreasing under CPTP maps. Since the set
of PPT preserving operations includes LOCC\ operations, $R_{\mathbf{D}}(  A\!:\!B)_\rho  $ is also
monotone under LOCC operations.

Finally, we define the \emph{generalized Rains
information of a quantum channel} as
\begin{equation}
R_{\mathbf{D}}(  \mathcal{N})  := \sup_{\rho_{RA}} R_{\mathbf{D}%
}(R\!:\!B)_{\omega}, 
\end{equation}
where $\omega_{RB} = \mathcal{N}_{A\rightarrow B}(  \rho_{RA}) \, $.

Another critical property of the above quantities has to do with the set
$\RainsSet(  A:B)  $\ of operators over which we are optimizing.
That is, all operators in this set satisfy the property given in Lemma~2 of
\cite{R99}, which we recall now:

\begin{lemma}
[Lemma 2 of \cite{R99}]\label{lem:overlap}Let $\tau_{AB}\in\RainsSet(
A\!:\!B)$. Then the overlap of $\tau_{AB}$ with any  maximally entangled
state $\Phi_{AB}$ of Schmidt rank $M$ is at most $1/M$, i.e.,
$\operatorname{Tr}\left\{  \Phi_{AB}\tau_{AB}\right\}  \leq\frac{1}{M}.$
\end{lemma}
\noindent The same is true for $\sigma_{AB}\in\ \operatorname{PPT}(  A\!:\!B)  $
simply because $\operatorname{PPT}(  A\!:\!B)  \subseteq\RainsSet%
(  A\!:\!B)  $. 

\subsection{Covariance of Quantum Channels}

Covariant quantum channels have symmetries which allow us to simplify the set
of states over which we need to optimize their generalized Rains information.
Let $G$ be a finite group, and for every $g\in G$, let $g\rightarrow
U_{A}(  g)  $ and $g\rightarrow V_{B}(  g)  $ be unitary
representations acting on the input and output spaces of the channel,
respectively. Then a\ quantum channel $\mathcal{N}_{A\rightarrow B}$\ is
covariant with respect to these representations if the following relation
holds for all input density operators $\rho_A \in \cS(A)$ and group elements $g\in G$:%
\begin{equation}
\mathcal{N}_{A\rightarrow B}\!\left(  U_{A}(  g)  \rho_A U_{A}^{\dag
}(  g)  \right)  =V_{B}(  g)  \mathcal{N}_{A\rightarrow
B}(  \rho_A )  V_{B}^{\dag}(  g)  .
\end{equation}
We then have the following proposition which allows us to restrict the form of
the input states needed to optimize the generalized Rains information of a
covariant channel:

\begin{proposition}
\label{prop:covariance}
Let $\mathcal{N}_{A\rightarrow B}$ be a covariant channel with group $G$ as above and let $\rho_{A} \in \cS(A)$, $\phi^{\rho}_{RA}$ a purification of $\rho_A$, and $\rho_{RB} = \mathcal{N}_{A\rightarrow B}(\phi^\rho_{RA})$. Let $\bar{\rho}_{A}$ be the group average of $\rho_{A}$, i.e.,%
\begin{equation}
\bar{\rho}_{A} = \frac{1}{\left\vert G\right\vert }\sum_{g}%
U_{A}(  g)  \rho_A U_{A}^{\dag}(  g)  ,
\end{equation}
and let $\phi^{\bar{\rho}}_{RA}$ be a
purification of $\bar{\rho}_A$ and $\bar{\rho}_{RB} = \mathcal{N}_{A\rightarrow B}(\phi^{\bar{\rho}}_{RA})$. Then,
$R_{\mathbf{D}}(R\!:\!B)_{\bar{\rho}} \geq R_{\mathbf{D}}(R\!:\!B)_{{\rho}}$
\end{proposition}

\begin{IEEEproof}
Given the purification $\phi^\rho_{RA}$, consider
the following state%
\begin{equation}
\left\vert \psi\right\rangle _{PRA}:=\sum_{g}\frac{1}{\sqrt{\left\vert
G\right\vert }}\vert g\rangle _{P}\left[  I_{R}\otimes U_{A}(
g)  \right]  \left\vert \phi^{\rho}\right\rangle _{RA}\text{.}%
\end{equation}
Observe that $\left\vert \psi\right\rangle _{PRA}$ is a purification of
$\overline{\rho}_{A}$ with purifying systems $P$ and $R$. Let $\tau_{PRB}$ be
an arbitrary operator in $\RainsSet(  PR\!:\!B)  $. Then the
chain of inequalities in \eqref{eq:long-chain-1}--\eqref{eq:long-chain-end} holds.%
\begin{figure*}
\begin{align}
&  \mathbf{D}(  \mathcal{N}_{A\rightarrow B}(\psi_{PRA})\Vert\tau
_{PRB}) \nonumber\\
& \qquad \geq\mathbf{D}\!\left(  \sum_{g}\frac{1}{\left\vert G\right\vert }\vert
g\rangle \langle g\vert _{P}\otimes\mathcal{N}_{A\rightarrow
B}(U_{A}(  g)  \phi_{RA}^{\rho}U_{A}^{\dag}(  g)
)\middle\Vert\sum_{g}p(  g)  \vert g\rangle \langle
g\vert _{P}\otimes\tau_{RB}^{g}\right) \label{eq:long-chain-1}\\
& \qquad =\mathbf{D}\!\left(  \sum_{g}\frac{1}{\left\vert G\right\vert }\vert
g\rangle \langle g\vert _{P}\otimes V_{B}(  g)
\mathcal{N}_{A\rightarrow B}(\phi_{RA}^{\rho})V_{B}^{\dag}(  g)
\middle\Vert\sum_{g}p(  g)  \vert g\rangle \langle
g\vert _{P}\otimes\tau_{RB}^{g}\right) \label{eq:classical-on-P}\\
& \qquad =\mathbf{D}\!\left(  \sum_{g}\frac{1}{\left\vert G\right\vert }\vert
g\rangle \langle g\vert _{P}\otimes\mathcal{N}_{A\rightarrow
B}(\phi_{RA}^{\rho})\middle\Vert\sum_{g}p(  g)  \vert g\rangle
\langle g\vert _{P}\otimes V_{B}^{\dag}(  g)  \tau
_{RB}^{g}V_{B}(  g)  \right) \\
& \qquad \geq\mathbf{D}\!\left(  \mathcal{N}_{A\rightarrow B}(\phi_{RA}^{\rho}%
)\middle\Vert\sum_{g}p(  g)  V_{B}^{\dag}(  g)  \tau_{RB}%
^{g}V_{B}(  g)  \right) \\
& \qquad \geq\inf_{\tau_{RB}\in\RainsSet(  R:B)  }\mathbf{D}(
\mathcal{N}_{A\rightarrow B}(\phi_{RA}^{\rho})\Vert\tau_{RB})   =R_{\mathbf{D}}(R\!:B)_{{\rho}} \,.
\label{eq:long-chain-end}
\end{align}
\end{figure*}
The first inequality follows from monotonicity of the generalized divergence
$\mathbf{D}$ under a dephasing of the $P$ register (where the dephasing
operation is given by $\sum_{g}\vert g\rangle \left\langle
g\right\vert \cdot\vert g\rangle \langle g\vert $). The
first equality follows from the assumption of channel covariance. The second
equality follows from invariance of the generalized divergence under
unitaries, with the unitary chosen to be%
\begin{equation}
\sum_{g}\vert g\rangle \langle g\vert _{P}\otimes
V_{B}^{\dag}(  g)  .
\end{equation}
Furthermore, this unitary does not take the state out of the class
$\RainsSet$, i.e.
\begin{equation}
\sum_{g}p(  g)  \vert g\rangle \langle g\vert
_{P}\otimes V_{B}^{\dag}(  g)  \tau_{RB}^{g}V_{B}(  g)
\in\RainsSet(  PR\!:\!B)  .
\end{equation}
This is because, in this case, one could also implement this operation as a
classically controlled LOCC\ operation, i.e., a von Neumann measurement
$\left\{  \vert g\rangle \langle g\vert \right\}  $ of
the register $P$ followed by a rotation $V_{B}^{\dag}(  g)  $ of
the $B$ register. One can do so here because both arguments to $\mathbf{D}$
in (\ref{eq:classical-on-P}) are classical on $P$. The second inequality
follows because the generalized divergence $\mathbf{D}$ is monotone under the
discarding of the register $P$. The final inequality results from taking a
minimization, and the final equality is by definition. Since $\tau_{PRB}$ is
chosen to be an arbitrary operator in $\RainsSet(  PR\!:\!B)  $, it
follows that%
\begin{equation}
\inf_{\tau_{PRB} \in \RainsSet(PR:B)} \mathbf{D}(  \mathcal{N}_{A\rightarrow B}(\psi_{PRA}) \| \tau_{PRB} ) \geq R_{\mathbf{D}}(R\!:\!B)_{{\rho}}
\end{equation}
The conclusion then follows because all purifications are related by an isometry
acting on the purifying system and the quantity $R_{\mathbf{D}}$ is invariant under
isometries acting on the purifying system.
\end{IEEEproof}

\subsection{Specializing to R\'enyi Divergence}

At this point we specialize our discussion to a particular type of R\'enyi divergence. 
For $\rho\in\mathcal{S}$,
$\sigma\in\mathcal{P}$ and $\alpha \in (0,1) \cup (1,\infty)$, we define the \emph{sandwiched R\'{e}nyi relative entropy}
of order $\alpha$ as~\cite{MDSFT13,WWY13}%
\begin{equation}
\widetilde{D}_{\alpha}(  \rho\Vert\sigma)  := 
\frac{1}{\alpha
-1}\log  \tr\left\{  \left(  \sigma^{(  1-\alpha)  /2\alpha}\rho
\sigma^{(  1-\alpha)  /2\alpha}\right)  ^{\alpha}\right\}
\end{equation}
if supp$(  \rho)  \subseteq\text{supp}(  \sigma)
\text{ or } \alpha\in(0,1)$ and it is equal to
$+\infty$ otherwise.
The sandwiched R\'enyi relative entropy is defined for $\alpha\in\{1,\infty\}$ by taking the respective limit. In particular,
  $\lim_{\alpha \to 1} \widetilde{D}_{\alpha}(\rho\|\sigma) = D(\rho\|\sigma)$~\cite{MDSFT13,WWY13}.
In the following, we restrict our attention to the regime for which $\alpha > 1$. 
The sandwiched R\'enyi relative entropy is monotone under CPTP\ maps $\mathcal{N}$ for such values of $\alpha$~\cite{FL13,B13monotone,MO13} and thus constitutes a generalized divergence as discussed above. In particular, we note that the trace term 
\begin{align}
  (\rho, \sigma) \mapsto \widetilde{Q}_{\alpha}(\rho\|\sigma) = \tr\left\{  \left(  \sigma^{(  1-\alpha)  /2\alpha}\rho
\sigma^{(  1-\alpha)  /2\alpha}\right)  ^{\alpha}\right\} \label{eq:traceterm}
\end{align}
is jointly convex in its arguments for $\alpha > 1$~\cite{FL13}. 

We denote the corresponding Rains relative entropy by $\widetilde{R}_{\alpha}(A\!:\!B)_{\rho}$, or simply use the shorthand $\widetilde{R}_{\alpha}(\rho_{AB}) = \widetilde{R}_{\alpha}(A\!:\!B)_{\rho}$ if the bipartition is evident.
The R\'enyi Rains relative entropy is \emph{subadditive}: namely, for any two bipartite states $\rho_{AB}$ and $\sigma_{A'B'}$ and $\alpha \in (0,1)\cup(1,\infty)$, we have
\begin{align}
\label{eq:sub-add}
  \widetilde{R}_{\alpha}(AA'\!:\!BB')_{\rho \otimes \sigma} \leq
  \widetilde{R}_{\alpha}(A\!:\!B)_{\rho} + \widetilde{R}_{\alpha}(A'\!:\!B')_{\sigma} .
\end{align}
This follows directly from the additivity of the sandwiched R\'enyi relative entropy with respect to tensor-product states and from the definition, since we can always restrict the minimization to product states.

 As a consequence of the joint convexity of the expression in~\eqref{eq:traceterm} for $\alpha > 1$ \cite[Proposition~3]{FL13}, its point-wise minimum over $\sigma$ is still a convex function of $\rho$. Moreover, taking into account the logarithm and prefactor in the definition of the R\'enyi divergence, we conclude that
the R\'enyi Rains relative entropy is \emph{quasi-convex}: i.e., if $\rho_{AB}$ decomposes as $\rho_{AB} = \int \mathrm{d}\mu(x) \rho_{AB}^x$ then
\begin{align}
\widetilde{R}_{\alpha}(\rho_{AB}) \leq \sup_{x} \widetilde{R}_{\alpha}(\rho_{AB}^x) \, .
\label{eq:quasi-convex}
\end{align}

The \emph{R\'enyi Rains information of a quantum channel} $\cN_{A\to B}$ is defined as
\begin{equation}
\widetilde{R}_{\alpha}(  \mathcal{N})  :=\sup_{\rho_{RA}}%
\inf_{\tau_{RB}\in\RainsSet(  R:B)  }\widetilde{D}_{\alpha
}(  \mathcal{N}_{A\rightarrow B}(  \rho_{RA})  \Vert\tau
_{RB})  .
\end{equation}
It suffices to
perform the maximization in $\widetilde{R}_{\alpha}(  \mathcal{N})$
over pure bipartite states $\rho_{RA}$, due to the quasi-convexity of
$\widetilde{R}_\alpha$ whenever $\alpha > 1$ (as discussed above).
As a result, it suffices for the dimension of the reference system $R$ to be no
larger than the dimension of the channel input $A$, due to the
well known Schmidt decomposition.

The R\'enyi Rains information converges to $R(
\mathcal{N})  $ in the limit as $\alpha$ approaches one from above.
This is shown in the following lemma, whose proof is provided in Appendix~\ref{sec:ogawa-nagaoka}.
\begin{lemma}
\label{lm:continuity}
  For any quantum channel $\mathcal{N}$ and $\alpha > \beta > 1$, we have
\begin{align}
  \widetilde{R}_{\alpha}(  \mathcal{N})
 & \geq \widetilde{R}_{\beta}(  \mathcal{N}) \geq R(  \mathcal{N}), \\
\lim_{\alpha\rightarrow1^{+}}\widetilde{R}_{\alpha}(  \mathcal{N})
& =R(  \mathcal{N})  .
\end{align}
\end{lemma}

\subsection{Relating Fidelity of an Entanglement Generation Code to
the R\'enyi Rains Information of a Channel}

The power of the generalized divergence framework is that it allows us to
relate rate and fidelity to an information quantity. The usual approach is to
compare the states resulting from any code to a set of states resulting from a
\textquotedblleft useless channel.\textquotedblright\ For the transmission of
classical information, the only set of useless channels are those which trace
out the input to the channel and replace it with an arbitrary density
operator, effectively \textquotedblleft cutting the communication
line.\textquotedblright\ However, for the transmission of quantum information,
there are more interesting classes of \textquotedblleft useless
channels\textquotedblright~\cite{SS12}. For example, it is well known that a
PPT\ entanglement binding channel has zero quantum capacity \cite{HHH00}. More
generally, the bound in Lemma~\ref{lem:overlap}\ establishes that if both the
input to the channel and the reference system are replaced with an operator
$\tau_{RB}\in\RainsSet(R\!:\!B)$, then the fidelity with a maximally entangled
state can never be larger than $1/M$. Since for a memoryless channel we are
taking $M=2^{nQ}$, this overlap will be exponentially small with the number of
channel uses, so that \textquotedblleft channels\textquotedblright\ that
replace with $\tau_{RB}$ cannot send any quantum information reliably.

The following proposition gives a \textquotedblleft one-shot\textquotedblright%
\ bound on the fidelity of any CPPP assisted EG code:

\begin{proposition}
\label{lem:one-shot} Let $\cN$ be a quantum channel. Any \CPP assisted EG code $\cC$ on $\cN$ obeys the following bound. For all $\alpha>1$,%
\begin{equation}
F(\cC, \cN) \leq2^{-\left(  \frac{\alpha-1}{\alpha}\right)  \left(  \log |\cC| -  \widetilde
{R}_{\alpha}\left(  \mathcal{N} \right)  \right)  } \, .
\label{eq:fidelity-bound-rains}%
\end{equation}
\end{proposition}

\begin{IEEEproof}
Let $\cC = (M, \cE_{A_0B_0\ \to \tilde{A}A\tilde{B}}, \cD_{\tilde{A}B\tilde{B} \to \hat{A}\hat{B}})$ as in~\eqref{eq:code} and recall the state $\rho_{\tilde{A}B\tilde{B}}$ in~\eqref{eq:state-output-from-channel} prior to decoding and the state $\omega_{\hat{A}\hat{B}} = \cD_{\tilde{A}B\tilde{B}\to\hat{A}\hat{B}}(\rho_{\tilde{A}B\tilde{B}})$. 
The binary test channel $\mathcal{B}_{\hat{A}\hat{B}\rightarrow Z}$ outputs a flag
indicating if the state is maximally entangled in the state $\Phi_{\hat{A}\hat{B}}$ or orthogonal to it:%
\begin{multline}
\mathcal{B}_{\hat{A}\hat{B}\rightarrow Z}(  \cdot)  :=\tr%
\left\{  \Phi_{\hat{A}\hat{B}}(  \cdot)  \right\}  \vert
1\rangle \langle 1\vert \\+\tr\left\{  \left(
I_{\hat{A}\hat{B}}-\Phi_{\hat{A}\hat{B}}\right)  (  \cdot)  \right\}
\vert 0\rangle \langle 0\vert .
\label{eq:entanglement-test}%
\end{multline}

Furthermore, consider an arbitrary subnormalized state $\tau_{\tilde{A}B\tilde{B}} \in \RainsSet(\tilde{A}\!:\!B\tilde{B})$ and observe that
$\mathcal{D}_{\tilde{A}B\tilde{B}\to\hat{A}\hat{B}}(\tau_{\tilde{A}B\tilde{B}})
\in \RainsSet(\hat{A}\!:\!\hat{B})$
because the decoding operator is restricted to be LOCC.
For ease of presentation, we set
\begin{align}
p & = \tr\left\{  \Phi_{\hat{A}\hat{B}}\, \mathcal{D}_{\tilde{A}B\tilde{B}\rightarrow\hat{A}\hat{B}} (  \tau_{\tilde{A}B\tilde{B}})  \right\}, \label{eq:overlap-decoded-PPT-12}\\
 F & = F(\cC,\cN) = \tr\left\{ \Phi_{\hat{A}\hat{B}}\, \omega_{\hat{A}\hat{B}} \right\} , \label{eq:overlap-decoded-PPT}%
\end{align}
and assume without loss of generality that the operator
$\tau_{\tilde{A}B\tilde{B}}$ is chosen such that
$p \in (0,\tr\left\{  \tau_{\tilde{A}B\tilde{B}} \right\})$ (otherwise, $\tau_{\tilde{A}B\tilde{B}}$ would not be a good choice because we would have
$\widetilde{D}_{\alpha}(  \rho_{\tilde{A}B\tilde{B}}\Vert\tau_{\tilde{A}B\tilde{B}}) = +\infty$).
Applying monotonicity of the divergence under the decoding map $\cD$ and the test $\cB$, we find
\begin{align}
& \widetilde{D}_{\alpha}(  \rho_{\tilde{A}B\tilde{B}}\Vert\tau_{\tilde{A}B\tilde{B}})   \nonumber\\
&
\geq\widetilde{D}_{\alpha}(  \mathcal{B}_{\hat{A}\hat{B}\rightarrow Z}(
\omega_{\hat{A}\hat{B}})  \Vert\mathcal{B}_{\hat{A}\hat{B}\rightarrow Z}(
\mathcal{D}_{\tilde{A}B\tilde{B}\rightarrow\hat{A}\hat{B}}(  \tau_{\tilde{A}B\tilde{B}})  )  )
\label{eq:renyi-develop-1}\\
&  =\frac{1}{\alpha-1}\log\left[  F^{\alpha}p^{1-\alpha}+\left(  1-F\right)
^{\alpha}\left[  \tr\left\{  \tau_{\tilde{A}B\tilde{B}} \right\}  -p\right]  ^{1-\alpha
}\right] \\
&  \geq\frac{1}{\alpha-1}\log\left[  F^{\alpha}p^{1-\alpha}\right]   \\
& \geq\frac{1}{\alpha-1}\log\left[  F^{\alpha}\left(  1/M\right)  ^{1-\alpha
}\right]  \\
& =\frac{\alpha}{\alpha-1}\log F+\log M \label{eq:renyi-develop-last} .%
\end{align}
The second inequality follows by discarding the second term $\left(
1-F\right)  ^{\alpha}\left[  \tr\left\{  \tau_{RB}\right\}  -p\right]
^{1-\alpha}$ (recall that we are considering $\alpha>1$). The third inequality
follows from (\ref{eq:overlap-decoded-PPT-12}) and Lemma~\ref{lem:overlap}.

Now, recall that the state $\rho_{\tilde{A}A\tilde{B}}$ is in $\Sepp(\tilde{A}A\!:\!\tilde{B})$ and can thus be decomposed as a convex combination of tensor products of pure states. Using the quasi-convexity of $\widetilde{D}_{\alpha}$ in the first argument, we find that there exist pure states $\sigma_{\tilde{A}A}$ and $\sigma_{\tilde{B}}$ such that for every
$\tau_{\tilde{A}B} \in \RainsSet(\tilde{A}:B)$, we have
\begin{align}
  & \!\!\!\!\widetilde{D}_{\alpha}( \cN_{A\to B} ( \sigma_{\tilde{A}A} )  \| \tau_{\tilde{A}B}) \nonumber \\
  &=
   \widetilde{D}_{\alpha}( \cN_{A\to B} ( \sigma_{\tilde{A}A} ) \otimes \sigma_{\tilde{B}} \| \tau_{\tilde{A}B} \otimes \sigma_{\tilde{B}} ) \\
  &\geq  \widetilde{D}_{\alpha}( \cN_{A\to B} ( \rho_{\tilde{A}A\tilde{B}} ) \| \tau_{\tilde{A}B} \otimes \sigma_{\tilde{B}} )\\
  & \geq \frac{\alpha}{\alpha-1}\log
F+\log M,
\end{align}
where in the last line we apply the development
in \eqref{eq:renyi-develop-1}-\eqref{eq:renyi-develop-last}
given that $\tau_{\tilde{A}B} \otimes \sigma_{\tilde{B}} \in
\RainsSet(  \tilde{A}\!:\!B\tilde{B}) $. Since the above bound holds for all
$\tau_{\tilde{A}B} \in \RainsSet(\tilde{A}:B)$, we can conclude that
\begin{multline}
\inf_{\tau_{\tilde{A}B} \in \RainsSet(\tilde{A}:B) }\widetilde{D}_{\alpha}( \cN_{A\to B} ( \sigma_{\tilde{A}A} )  \| \tau_{\tilde{A}B}) 
\\\geq \frac{\alpha}{\alpha-1}\log
F+\log M.
\end{multline}
We can finally remove the dependence on any particular code by optimizing over
all inputs to the channel. Identifying $\tilde{A}$ with $R$ to simplify notation, we find
\begin{align}
\widetilde{R}_{\alpha}(  \mathcal{N})  & =\sup_{\rho_{RA}}\inf
_{\tau_{RB}\in\RainsSet(  R:B)  }\widetilde{D}_{\alpha}(
\mathcal{N}_{A\rightarrow B}(  \rho_{RA})  \Vert\tau_{RB})\\
& \geq\frac{\alpha}{\alpha-1}\log F+\log M.
\end{align}
This bound is then equivalent to (\ref{eq:fidelity-bound-rains}).
\end{IEEEproof}

\section{Weak Subadditivity of the $\alpha$-Rains Information for Memoryless
Channels}
\label{sec:weaksub}

In this section, we prove an important theorem, which is critical for
concluding that the Rains information of a channel is a strong converse rate
for quantum communication.
Before we commence, we need the following technical property.

\begin{lemma}
\label{lem:mono-inc} Let $\alpha > 1$, $\rho, \rho' 
\in \mathcal{S}$ and $\sigma \in \mathcal{P}$. If $\rho \leq \gamma \rho'$ for some $\gamma \geq 1$, then
\begin{equation}
\widetilde{D}_{\alpha} (  \rho\Vert\sigma)  \leq \frac{\alpha}{\alpha-1} \log \gamma + \widetilde{D}_{\alpha}(  \rho'\Vert\sigma) \,.
\end{equation}
\end{lemma}

\begin{IEEEproof}
From the assumption that $\rho\leq \gamma \rho^{\prime}$, we get $\sigma^{(
1-\alpha)  /2\alpha}\rho\sigma^{(  1-\alpha)  /2\alpha}%
\leq \gamma \sigma^{(  1-\alpha)  /2\alpha}\rho^{\prime}\sigma^{(
1-\alpha)  /2\alpha}$. Then we have that
\begin{multline}
\tr \{ (\sigma^{(
1-\alpha)  /2\alpha}\rho\sigma^{(  1-\alpha)  /2\alpha})^\alpha\}
\\\leq
\gamma^\alpha \tr \{ (\sigma^{(  1-\alpha)  /2\alpha}\rho^{\prime}\sigma^{(
1-\alpha)  /2\alpha})^\alpha \}
\end{multline}
because
tr$\left\{  f(  P)  \right\}  \leq\ $tr$\left\{  f(  Q)
\right\}  $ for $P\leq Q$ and $f$ a monotone increasing function (see, e.g.,
\cite[Lemma~III.6]{MO13}). Taking logarithms and dividing by $\alpha -1$ gives the statement of the lemma.
\end{IEEEproof}

We are ready to prove that the Rains information of the channel obeys a weak subadditivity property.

\begin{theorem}
\label{thm:weak-subadditivity}
Let $\cN_{A\to B}$ be a quantum channel. For all $\alpha>1$ and $n \in \mathbb{N}$, we have
\begin{equation}
\widetilde{R}_{\alpha}(  \mathcal{N}^{\otimes n})  \leq
n\widetilde{R}_{\alpha}(  \mathcal{N})  +\frac{\alpha\left\vert
A\right\vert ^{2}}{\alpha-1}\log n \label{eq:weak-subadd} .
\end{equation}
\end{theorem}

\begin{IEEEproof}
To begin with, we observe that a tensor-power channel is covariant with
respect to permutations of the input and output systems, in the sense that%
\begin{multline}
\forall\pi\in S_{n}:W_{B^{n}}^{\pi}\mathcal{N}^{\otimes n}(  \rho_{A^{n}%
})  \left(  W_{B^{n}}^{\pi}\right)  ^{\dag}\\=\mathcal{N}^{\otimes
n}\left(  W_{A^{n}}^{\pi}\rho_{A^{n}}\left(  W_{A^{n}}^{\pi}\right)  ^{\dag
}\right)  ,
\end{multline}
where $W_{A^{n}}^{\pi}$ and $W_{B^{n}}^{\pi}$ are unitary representations of
the permutation $\pi$, acting on the input space$~A^{n}$ and the output space
$B^{n}$, respectively. So, letting $\left\vert \phi^{\rho}\right\rangle
_{RA^{n}}$ denote a purification of $\rho_{A^{n}}$, we can apply
Proposition~\ref{prop:covariance}\ to find that
\begin{equation}
\widetilde{R}_{\alpha}(  \mathcal{N}^{\otimes n}(
\phi_{RA^{n}}^{\rho})  )  \leq\widetilde{R}_{\alpha}(
\mathcal{N}^{\otimes n}(  \phi_{RA^{n}}^{\overline{\rho}%
})  )  , \label{eq:1st-proof-line}%
\end{equation}
where $\phi_{RA^{n}}^{\overline{\rho}}$ is a purification of the permutation
invariant state $\overline{\rho}_{A^{n}}$.
Now, this purification $\phi_{RA^{n}}^{\overline{\rho}}$ is related by a
unitary on the reference system $R$ to a state
$\left\vert \psi\right\rangle _{\hat{A}^{n}A^{n}}\in\text{Sym}((\hat{A}\otimes
A)^{\otimes n})$,
where $\hat{A}\simeq A$~\cite[Lemma 4.3.1]{RennerThesis}. So it follows that%
\begin{align}
\widetilde{R}_{\alpha}(  \mathcal{N}^{\otimes n}(
\phi_{RA^{n}}^{\rho})  )  & \leq\widetilde{R}_{\alpha}(
\mathcal{N}^{\otimes n}(  \phi_{RA^{n}}^{\overline{\rho}%
})  )  \nonumber \\
& =\widetilde{R}_{\alpha}(  \mathcal{N}^{\otimes n}(  \psi_{\hat{A}^{n}A^{n}})  )  .
\label{eq:2nd-proof-line}%
\end{align}
For such a state in Sym$((\hat{A}\otimes A)^{\otimes n})$, we observe (see, e.g., \cite{CKR09}) that
\begin{equation}
\psi_{\hat{A}^{n}A^{n}}\leq n^{\left\vert A\right\vert ^{2}}\omega_{\hat{A}^{n}A^{n}}^{(n)  }, 
\quad \textrm{where} \quad
\omega_{\hat{A}^{n}A^{n}}^{\left(  n\right)  } := \int d\mu(  \varphi)
\, \varphi_{\hat{A}A}^{\otimes n}, \label{eq:de-finetti-state} 
\end{equation}
with $\mu(  \varphi)  $ denoting the uniform probability measure on
the unit sphere consisting of pure bipartite states~$\varphi
_{\hat{A}A}$. 

Employing Lemma~\ref{lem:mono-inc}, we find
that%
\begin{align}
  \widetilde{R}_{\alpha}(  \mathcal{N}^{\otimes n}(
\psi_{\hat{A}^{n}A^{n}})  ) \leq \frac{\alpha\left\vert A\right\vert ^{2}}{\alpha-1}\log n + \widetilde{R}_{\alpha}(  \mathcal{N}^{\otimes n}(
\omega_{\hat{A}^{n}A^{n} })  ) .
\end{align}
and since the right hand side does not depend on the state $\rho_{A^n}$ anymore, this yields
\begin{align}
  \widetilde{R}_{\alpha}(\cN^{\otimes n}) & \leq \frac{\alpha\left\vert A\right\vert ^{2}}{\alpha-1}\log n + \widetilde{R}_{\alpha}(  \mathcal{N}^{\otimes n}(
\omega_{\hat{A}^{n}A^{n} })  ) \\
& \leq \frac{\alpha\left\vert A\right\vert ^{2}}{\alpha-1}\log n + \sup_{\phi_{\hat{A}A}} \widetilde{R}_{\alpha}(  \mathcal{N}^{\otimes n}(
\phi_{\hat{A}A}^{\otimes n} )  ) \label{eq:use-this}
\end{align}
where we used the quasi-convexity of $\widetilde{R}_{\alpha}$ in the state (cf.~\eqref{eq:quasi-convex}) and the definition of $\omega_{\hat{A}^{n}A^{n}}^{(n)}$ in~\eqref{eq:de-finetti-state} to establish the second inequality. Finally, $\widetilde{R}_{\alpha}$ is subadditive for product states as seen in~\eqref{eq:sub-add}, and thus
\begin{align}
\widetilde{R}_{\alpha}\big(  \mathcal{N}^{\otimes n}(
\phi_{\hat{A}A}^{\otimes n} )  \big) \leq n \widetilde{R}_{\alpha}(  \mathcal{N}(\phi_{\hat{A}A})  ).
\end{align}
Combining this with~\eqref{eq:use-this} concludes the proof.
\end{IEEEproof}

A consequence of Theorem~\ref{thm:weak-subadditivity} is that the
Rains information of a channel is weakly subadditive. This corollary is
required in order to set some of the claims in \cite{SS08,SSW08} on a firm
foundation. We provide its proof in Appendix~\ref{sec:appendix}.

\begin{corollary}
\label{cor:von-Neumann-subadd} The Rains information of a quantum channel is
weakly subadditive, in the sense that%
\begin{equation}
\limsup_{n \to\infty} \frac{1}{n}R(  \mathcal{N}^{\otimes n})  \leq
R(  \mathcal{N})  .
\end{equation}

\end{corollary}

\section{The Rains Information is a Strong Converse Rate for
Quantum Communication}
\label{sec:main}

We are ready to state our main result, which is that the Rains information of the channel is an upper bound on the strong converse \CPP assisted quantum capacity.

\begin{theorem}
\label{thm:strong-converse}
  For any quantum channel $\cN$, we have $Q_{\CPPsub}^{\dagger}(\cN) \leq \inf_{\ell \in \mathbb{N}} \frac{1}{\ell} R(\cN^{\otimes \ell})$.
\end{theorem}

Before we prove this result, we first state a technical proposition which implies that the fidelity decreases exponentially fast as the number of channel uses increases, with the exponent bounded in~\eqref{eq:exponent} below. (However, we do not know if the exponent in~\eqref{eq:exponent} is optimal.)

\begin{proposition}
\label{pr:strong-converse}
Let $\cN$ be a quantum channel. Consider any sequence of codes $\{ \cC_n \}_{n \in \mathbb{N}}$, where $\cC_n$ is a \CPP assisted EG code for $\cN^{\otimes n}$. Then the rate of this sequence, $r = \liminf_{n\to\infty} \frac{1}{n} \log |\cC_n|$, satisfies 
\begin{align}
   \liminf_{n\to\infty} \left\{ - \frac{1}{n} \log F(\cC_n,\cN^{\otimes n}) \right\} \geq \sup_{\alpha > 1} \left\{ \frac{\alpha-1}{\alpha} \big( r - \widetilde{R}_{\alpha}(\cN) \big) \right\}. \label{eq:exponent}
\end{align}
\end{proposition}

\begin{IEEEproof}
First, by definition of $r$, for any $\delta > 0$, there exists an $N_0 \in \mathbb{N}$ such that $\frac{1}{n} \log |\cC_n| \geq r - \delta$ for all $n \geq N_0$. For such $n$ and any $\alpha > 1$, we employ Proposition~\ref{lem:one-shot} to find the following bound:
\begin{align}
F\left(\cC_n,\cN^{\otimes n}\right) &\leq2^{-\left(  \frac{\alpha-1}{\alpha}\right)  \left(  \log
|\cC_n|-\widetilde{R}_{\alpha}(  \mathcal{N}^{\otimes n})  \right)  }\\
&\leq 2^{-\left(  \frac{\alpha-1}{\alpha}\right)  \left(  n(r-\delta) -\widetilde
{R}_{\alpha}(  \mathcal{N}^{\otimes n})  \right)  }\\
&  \leq2^{-\big(  \frac{\alpha-1}{\alpha}\big)  \big(  n(r -\delta) - n 
\widetilde{R}_{\alpha}(\mathcal{N})  -\frac{\alpha\left\vert
A\right\vert^{2}}{\alpha-1}\log n \big)  } \label{eq:use-weak}   
\\
& = n^{\left\vert A\right\vert^{2}}2^{-n\left(  \frac{\alpha-1}{\alpha
}\right)  \left(  r-\delta-\widetilde{R}_{\alpha}(  \mathcal{N}%
)  \right)  } .%
\end{align}
Here, we used the weak subadditivity result from Theorem~\ref{thm:weak-subadditivity} to establish~\eqref{eq:use-weak}. Hence, we find
\begin{align}
   \liminf_{n\to\infty} \left\{ - \frac{1}{n} \log F(\cC_n,\cN^{\otimes n}) \right\} \geq \frac{\alpha-1}{\alpha} \big( r - \delta - \widetilde{R}_{\alpha}(\cN) \big) 
\end{align}
and the statement of the theorem follows since we can choose $\delta$ and $\alpha$ arbitrarily.
\end{IEEEproof}

We briefly sketch a proof of the main theorem here for the case in which we want to show that $Q_{\CPPsub}^\dag(\cN) \leq R(\cN)$ and provide the full proof below. Consider any code with $r$ defined as in Proposition~\ref{pr:strong-converse}. If $r > R(\cN)$, then
by continuity of $\widetilde{R}_{\alpha}(\cN)$ as $\alpha \to 1^+$, there always exists an $\alpha > 1$ such that $r > \widetilde{R}_{\alpha}(\cN)$ as well. Thus, the right hand side of~\eqref{eq:exponent} is strictly positive and the fidelity thus vanishes.

\begin{IEEEproof}[Proof of Theorem~\ref{thm:strong-converse}]
Consider any sequence of codes $\{ \cC_n \}_{n \in \mathbb{N}}$, where $\cC_n$ is a \CPP assisted EG code for $\cN^{\otimes n}$ such that $r = \liminf_{n\to\infty} \frac{1}{n} \log |\cC_n| > \inf_{\ell \in \mathbb{N}} \frac{1}{\ell} R(\cN^{\otimes \ell})$. 
Then, by definition of the infimum there exists a value $\ell \in \mathbb{N}$ such that $\ell r > R(\cN^{\otimes \ell})$. Furthermore, by continuity of $\widetilde{R}_{\alpha}(\mathcal{N}^{\otimes \ell})$ as $\alpha \to 1^+$ (cf.~Lemma~\ref{lm:continuity}), there exists an $\alpha > 1$ such that $\ell r > \widetilde{R}_{\alpha}(\cN^{\otimes \ell})$.

Next, for any $j \in \{1, 2, \ldots, \ell\}$, consider the subsequence of codes $\{ \cC_{k\ell+j} \}_{k \in \mathbb{N}}$ and their embeddings as codes for the channels $\cN^{\otimes (k+1)\ell} =\big(\cN^{\otimes \ell}\big)^{\otimes (k+1)}$. (These codes are defined for $k\ell+j$ channels, and their embeddings will simply ignore the last $\ell-j$ channels.)
Proposition~\ref{pr:strong-converse} applied to the channel $\cN^{\otimes \ell}$ yields
 \begin{align}
&   \liminf_{k\to\infty} \left\{ - \frac{1}{k+1} \log F(\cC_{k\ell+j},\cN^{\otimes (k+1)\ell}) \right\} \nonumber\\
&\geq \frac{\alpha-1}{\alpha } \big( \ell r - \widetilde{R}_{\alpha}(\cN^{\otimes \ell}) \big) =: c > 0 .  \label{eq:nonzero-lower}
 \end{align}
 Here we used that $\frac{1}{\ell} \liminf_{k\to\infty} \frac{1}{k+1} \log |\cC_{k\ell+j}| = \liminf_{k\to\infty} \frac{1}{k\ell+j} \log |\cC_{k\ell+j}| - \frac{\ell - j}{(k\ell+l)(k\ell+j)} \log |\cC_{k\ell+j}|
 \geq r$ since the $\liminf$ of the subsequence $\{ \frac{1}{k\ell+j} \log |\cC_{k\ell+j}| \}_{k}$ is lower bounded by the $\liminf$ of the sequence $\{ \frac{1}{n} \log |\cC_n| \}_{n}$ and the second term vanishes. Moreover, \eqref{eq:nonzero-lower} yields
 \begin{align}
   & \liminf_{k\to\infty} \left\{ - \frac{1}{k\ell + j} \log F(\cC_{k\ell+j},\cN^{\otimes (k\ell + j}) \right\} \nonumber \\&=
   \liminf_{k\to\infty} \left\{ - \frac{1}{k\ell + j} \log F(\cC_{k\ell+j},\cN^{\otimes (k+1)\ell}) \right\} \\
   &\geq \liminf_{k\to\infty} \left\{ - \frac{1}{k\ell + \ell} \log F(\cC_{k\ell+j},\cN^{\otimes (k+1)\ell}) \right\}\\
   &\geq \frac{c}{\ell} \,.
 \end{align}
 Since this holds for all $j$, we conclude
 \begin{align}
 &   \liminf_{n\to\infty} \left\{ - \frac{1}{n} \log F(\cC_n,\cN^{\otimes n}) \right\} \nonumber \\
 & = \min_{j \in \{0,1, \dots \ell-1\}} \liminf_{k\to\infty} \left\{ - \frac{1}{k \ell+j} \log F(\cC_{k\ell+j},\cN^{\otimes k\ell+j}) \right\} \nonumber \\
&   \geq \frac{c}{\ell} \,.
 \end{align}
Hence, we conclude that the fidelity vanishes (exponentially fast in $n$) and that $\inf_{\ell \in \mathbb{N}} \frac{1}{\ell} R(\cN^{\otimes \ell})$ is a strong converse rate.
\end{IEEEproof}

Theorem~\ref{thm:strong-converse}\ establishes the Rains information of a
quantum channel as an upper bound on the strong converse capacity $Q_{\CPPsub}^{\dagger}(\cN)$ for \CPP assisted quantum communication over a channel $\cN$. Thus, in summary, the following inequalities hold for all quantum channels: \begin{multline}
I_{\rm c}(\cN) \leq \lim_{\ell \to \infty} \frac{1}{\ell} I_{\rm c}(\cN^{\otimes \ell})  = Q(\cN) \leq Q^{\dagger}(\cN) \\ \leq Q_{\CPPsub}^{\dagger}(\cN) \leq 
\inf_{\ell\geq 1} \frac{1}{\ell} R(\cN^{\otimes \ell})
\leq R(\cN) . \label{eq:hier}
\end{multline}

\section{Strong Converse Property for Quantum Communication over Dephasing Channels}
\label{sec:dephasing}

In this
section, we show that the Rains information of a generalized dephasing
channel~\cite{DS05,YHD05MQAC,itit2008hsieh,BHTW10} (also known as
\textquotedblleft Hadamard diagonal\textquotedblright\ channels \cite{KMNR07}
and \textquotedblleft Schur multiplier\textquotedblright\ channels
\cite{MW13}) is equal to the coherent
information of this channel. This extends \cite[Theorem~6.2]{R01} from ``maximally correlated states'' to generalized dephasing channels.
Consequently, the hierarchy in~\eqref{eq:hier} collapses for generalized dephasing channels.
In particular, we establish that the quantum capacity of
this class of channels obeys the strong converse property
and also that classical pre- and post-processing does not increase the capacity for these channels.

A generalized dephasing channel is any channel with an isometric extension of the form
\begin{equation}
U_{A\rightarrow BE}^{\mathcal{N}}:=\sum_{x=0}^{d-1}\left\vert
x\right\rangle _{B}\langle x\vert _{A}\otimes\left\vert \psi
_{x}\right\rangle _{E}, \label{eq:deph}
\end{equation}
where the states $\left\vert \psi_{x}\right\rangle $ are arbitrary (not
necessarily orthonormal). Note that these channels are degradable \cite{DS05}. 

\begin{proposition}
\label{prop:I_c=R-for-H}Let $\mathcal{N}$ be a generalized dephasing
channel of the form~\eqref{eq:deph}. Then
$I_{c}(  \mathcal{N})  =R(  \mathcal{N})$.
\end{proposition}

\begin{IEEEproof}
We have already seen in \eqref{eq:hier}
that $I_{c}(  \mathcal{N})  \leq R(  \mathcal{N})$ holds for all channels (recall also that this follows from 
\cite[Theorem~4]{PVP00}).
We now establish that the opposite inequality holds for a generalized dephasing
channel $\mathcal{N}$. Consider that any generalized dephasing channel $\mathcal{N}$\ obeys the
following covariance property:\footnote{The covariance in (\ref{eq:covariance-dephasing}) in fact holds for any
operators of the form $\sum_{x=0}^{d-1}\exp\left\{  i\varphi_{x}\right\}
\vert x\rangle \langle x\vert _{A}$ and $\sum
_{x=0}^{d-1}\exp\left\{  i\varphi_{x}\right\}  \vert x\rangle
\langle x\vert _{B}$ with $\varphi_{x}\in\mathbb{R}$, but it
suffices to consider only the operators in (\ref{eq:phase-op-B}) for our proof here.}
\begin{equation}
\mathcal{N}\left(  Z_{A}(  z)  \rho Z_{A}^{\dag}(
z)  \right)  =Z_{B}(  z)  \mathcal{N}(  \rho)
Z_{B}^{\dag}(  z)  , \label{eq:covariance-dephasing}%
\end{equation}
for $z\in\left\{  0,\cdots,d-1\right\}  $, where%
\begin{align}
Z_{A}(  z)  \vert x\rangle _{A}    & =\exp\left\{  2\pi
ixz/d\right\}  \vert x\rangle _{A},
\\
Z_{B}(  z)  \vert x\rangle _{B}    & =\exp\left\{  2\pi
ixz/d\right\}  \vert x\rangle _{B}. \label{eq:phase-op-B}%
\end{align}
 Furthermore, a uniform mixing of
these operators is equivalent to a \textquotedblleft completely
dephasing\textquotedblright\ channel:%
\begin{equation}
\frac{1}{d}\sum_{z=0}^{d-1}Z_{A}(  z)  (  \cdot)
Z_{A}^{\dag}(  z)  =\sum_{x=0}^{d-1}\vert x\rangle
\langle x\vert _{A}(  \cdot)  \vert x\rangle
\langle x\vert _{A},
\end{equation}
with the same true for the operators $\left\{  Z_{B}\left(
z\right)  \right\}  _{z\in\left\{  0,\ldots,d-1\right\}  }$. Then we can apply
Proposition~\ref{prop:covariance}\ to conclude that the Rains information of a
generalized dephasing channel is maximized by a state with a Schmidt
decomposition of the following form:%
\begin{equation}
\left\vert \varphi^{p}\right\rangle _{RA}:=\sum_{x}\sqrt{p_{X}(
x)  }\vert x\rangle _{R}\vert x\rangle _{A},
\label{eq:aligned-schmidt}%
\end{equation}
for some probability distribution $p_{X}(  x)  $ and some
orthonormal basis $\left\{  \vert x\rangle _{R}\right\}  $ for the
reference system $R$ (with the key result being that the basis $\left\{
\vert x\rangle _{A}\right\}  $ is \textquotedblleft aligned
with\textquotedblright\ the basis of the channel). That is,%
\begin{equation}
R(  \mathcal{N})  =\sup_{p_X}\inf_{\tau_{RB}%
\in\RainsSet(  R:B)  }D(  \mathcal{N}  (  \varphi_{RA}^{p})  \Vert\tau
_{RB})  . \label{eq:rains-hadamard}%
\end{equation}
Let $\Delta_{P}$ be a CPTP\ map constructed
as follows:%
\begin{align}
\Delta_{P}(  \cdot)   & = P(  \cdot)  P+\left(
I-P\right)  (  \cdot)  \left(  I-P\right)  , \\ \textrm{with} \qquad P  & =\sum_{x}\vert x\rangle \langle x\vert
_{R}\otimes\vert x\rangle \langle x\vert _{B} \,.
\end{align}
Then the following chain of inequalities holds:%
\begin{align}
I_{c}(  \mathcal{N})   &=\sup_{\varphi_{RA}}\inf_{\sigma_{B}%
}D(  \mathcal{N}  (
\varphi_{RA})  \Vert I_{R}\otimes\sigma_{B}) \\
&  \geq\sup_{p_X}\inf_{\sigma_{B}}D( \mathcal{N} (  \varphi_{RA}^{p})  \Vert
I_{R}\otimes\sigma_{B}) \\
&  \geq\sup_{p_X}\inf_{\sigma_{B}}D(  \Delta_{P}(
\mathcal{N}  (  \varphi_{RA}%
^{p})  )  \Vert\Delta_{P}(  I_{R}\otimes\sigma_{B})
) \\
&  =\sup_{p_X}\inf_{\sigma_{B}}
 D(  P(  \mathcal{N}  (  \varphi_{RA}^{p})
)  P\Vert P(  I_{R}\otimes\sigma_{B})  P) \\
&  =\sup_{p_X}\inf_{q}D(  \mathcal{N}  (  \varphi_{RA}^{p})  \Vert\sum
_{x}q(  x)  \vert x\rangle \langle x\vert
_{R}\otimes\vert x\rangle \langle x\vert _{B}) \\
&  \geq\sup_{p_X}\inf_{\tau_{RB}\in\RainsSet(  R:B)
}D(  \mathcal{N} (
\varphi_{RA}^{p})  \Vert\tau_{RB})  =R(  \mathcal{N})  .
\end{align}
The first inequality follows by
restricting the maximization to be over pure bipartite vectors of the form in
(\ref{eq:aligned-schmidt}). The second inequality follows from monotonicity of
the relative entropy under the CPTP\ map $\Delta_{P}$. The second equality
follows because%
\begin{multline}
D(  \Delta_{P}[  (  \text{id}_{R}\otimes\mathcal{N})
(  \varphi_{RA}^{p})  ]  \Vert\Delta_{P}(  I_{R}%
\otimes\sigma_{B})  )  \\=D(  P[  (  \text{id}%
_{R}\otimes\mathcal{N})  (  \varphi_{RA}^{p})  ]
P\Vert P(  I_{R}\otimes\sigma_{B})  P) \\
+D(  (  I-P)  [  (  \text{id}_{R}\otimes
\mathcal{N})  (  \varphi_{RA}^{p})  ]  (
I-P)  \Vert(  I-P)  (  I_{R}\otimes\sigma_{B})
(  I-P)  )
\end{multline}
since $P\perp I-P$, and because the state $\left(  \text{id}_{R}%
\otimes\mathcal{N}\right)  (  \varphi_{RA}^{p})  $ has no
support in the subspace onto which $I-P$ projects, so that the last term above is equal to zero.
%
The third equality follows because
\begin{align}
P\left[  \left(  \text{id}_{R}\otimes\mathcal{N}\right)  \left(
\varphi_{RA}^{p}\right)  \right]  P & =\left(  \text{id}_{R}\otimes
\mathcal{N}\right)  (  \varphi_{RA}^{p}), \\
P\left(  I_{R}\otimes\sigma_{B}\right)  P & =\sum_{x}q(  x)
\vert x\rangle \langle x\vert _{R}\otimes\left\vert
x\right\rangle \langle x\vert _{B},
\end{align}
for some distribution $q(  x)  =\langle x\vert
_{B}\sigma_{B}\vert x\rangle _{B}$. The last inequality follows
because the state
$\sum_{x}q(  x)  \vert x\rangle \langle x\vert
_{R}\otimes\vert x\rangle \langle x\vert _{B}%
\in\RainsSet(  R\!:\!B)$,
and the final equality follows from (\ref{eq:rains-hadamard}).
\end{IEEEproof}

\section{Strong Converse for Classical Communication\\Assisted Capacity of Erasure Channels}
\label{sec:erasure}

In this section, we consider a class of erasure channels of the form
\begin{align}
  \cN_{A \to B} : \rho_A \mapsto (1-p) \rho_{B} + p | e \rangle\!\langle e|_{B}, \label{eq:erasure}
\end{align} 
where $p \in [0,1]$ is the erasure probability,
$\rho_B$ is an isometric embedding of $\rho_A$ into $B$, and
$\vert e\rangle$ is a quantum state orthogonal to $\rho_{B}$.

Using the notion of entanglement cost of a channel, it was shown in \cite{BBCW13} that the strong converse two-way assisted quantum capacity of the erasure channel 
is equal to $(1-p) \log |A|$. An alternate way of proving this (implicit in prior work) is as follows:
\begin{enumerate}
\item Use the fact that the erasure channel can be realized by the action of teleportation on the state $\cN_{A \to B}(\Phi_{RA})$. As a consequence, any two-way assisted protocol for quantum communication can be simulated by an entanglement distillation protocol for the state $\cN_{A \to B}(\Phi_{RA})$. 
The authors of \cite[Section~V]{BDSW96} devised this teleportation simulation argument for discrete-variable channels,
and it was subsequently generalized to more channels, including continuous-variable channels, in \cite{PLOB15}.
\item Apply the strong converse theorem from \cite{H06} for distillable entanglement of a state and evaluate the Rains relative entropy.\footnote{Note here that one could also apply the relative entropy of entanglement
to obtain an upper bound
as done in \cite{PLOB15}, but this bound would be generally weaker than that given by the Rains relative entropy.}
\end{enumerate}
This latter approach is similar in spirit to what we do below, which however utilizes the methods given in this paper.

\begin{proposition}
  Let $\cN$ be an erasure channel of the form~\eqref{eq:erasure}. Then, $Q_{\CPPsub}(\cN) = Q_{\CPPsub}^{\dagger}(\cN) = (1-p) \log |A|$.
\end{proposition}

\begin{IEEEproof}
First note that erasure channels are covariant under the discrete Heisenberg--Weyl unitary group acting on $A$, and thus Proposition~\ref{prop:covariance} implies that
\begin{align}
  R(\cN) & = R(\cN(\psi_{RA})) \\
  & \leq D \big( \cN(\psi_{RA}) \big\| \tau_{RB} \big)
  \\
  & = H(\tau_{RB}) - H(\rho_{RB})
\end{align}
where $\psi_{RA}$ is a maximally entangled state, $\rho_{RB} = \cN(\psi_{RA})$, and $\tau_{RB}$ is chosen as follows:
\begin{align}
  |\psi\rangle_{RA} & = \sum_x \frac{1}{\sqrt{|A|}} |x\rangle_R \otimes |x\rangle_A , \\ 
  \tau_{RB} & = \sum_x \frac{1}{|A|} |x\rangle\!\langle x|_R \otimes \cN\big(|x\rangle\!\langle x|_A\big)  \,.
\end{align}
Evaluating this, we find
\begin{align}
H(RB)_\tau & = H(R)_\tau + H(B|R)_\tau \\
 & = \log |A| + \sum_x \frac{1}{|A|} H(B)_{\cN(|x\rangle\!\langle x|_A)} \\
 & = \log |A| + \sum_x \frac{1}{|A|} [H((1-p)|x\rangle\!\langle x|_A + p |e\rangle\langle e|)] \\
 & = \log |A| + h_2(p) ,\\
H(RB)_\rho & = H( (1-p) \psi_{RA} + p \pi_R \otimes |e\rangle \langle e |) \\
& = h_2(p) + (1-p) H( \psi_{RA}) + p H(\pi_R \otimes |e\rangle \langle e |) \\
& = h_2(p) + p \log|A| .
\end{align}
In the above, $H(B|R) := H(BR) - H(R)$ is the conditional entropy.
Therefore, we conclude that $Q_{\CPPsub}^{\dagger}(\cN) \leq R(\cN) \leq (1-p) \log |A|.$

It is well known that any rate $r < (1-p) \log |A|$ can be achieved with \CPP assistance
\cite{PhysRevLett.78.3217}. We review this argument briefly. Let Alice prepare and send $n$ maximally entangled states through $\cN^{\otimes n}$. Bob then records which instances of the channel led to an erasure, and communicates this to Alice. They will then use the correctly transmitted states to distill a maximally entangled state of dimension $r n$. This will succeed whenever $r n / \log |A|$ is smaller than the number of correctly transmitted states, which will happen with probability 1 as $n \to \infty$ as long as $r < (1-p) \log |A|$. This establishes that $Q_{\CPPsub}(\cN) \geq (1-p) \log |A|$ and concludes the proof.
\end{IEEEproof}

Finally, recall that the erasure channel is degradable for $p\leq 1/2$ \cite{DS05}, and we can thus calculate its unassisted quantum capacity to be $Q(\cN) = \max\{(1-2p) \log |A|, 0\}$~\cite{PhysRevLett.78.3217}. Thus,
we have an example of a channel where $Q(\cN) < R(\cN)$, which means that our upper bound in terms of the Rains information of the channel is not sufficient to establish the strong converse property for general degradable channels.

\section{Conclusion}

\label{sec:conclusion}This paper has established that the Rains information of
a quantum channel is a strong converse rate for quantum communication. The main application
of the first result is to establish the strong converse property for the quantum
capacity of all generalized dephasing channels.
Going forward from here, there are several questions to consider. First, are
there any other channels besides the generalized dephasing ones for which the
Rains information is equal to the coherent information?\ If true, the theorems
established here would establish the strong converse property for these channels. For
example, can we prove a strong converse theorem for the quantum capacity of
general Hadamard channels? 

Is it possible to show weak subadditivity of a R\'enyi coherent information quantity
for some class of channels, addressing the original question posed in \cite{SW12}? 
To this end, the developments in \cite{TBH13} might be helpful. 
Next, is it
possible to show that the Rains information of a general quantum channel represents a
strong converse rate for quantum communication assisted by interactive forward and backward classical communication? 
Similarly, can one show that the squashed
entanglement of a channel \cite{TGW14}\ is a strong converse rate for this task? 
Recent work has proved that the squashed entanglement of a quantum channel
is an upper bound on the quantum capacity with interactive forward and backward classical communication, and it could be that the quantities
defined in \cite{BSW14}\ would be helpful for settling this question. 

Finally, now that the strong converse holds for the
classical capacity, the quantum capacity, and the entanglement-assisted
capacity of all generalized dephasing channels, can we establish that the
strong converse holds for trade-off capacities of these channels, in the sense
of \cite{BHTW10,WH10b}? Can we establish second-order characterizations for
this class of channels, in the sense of \cite{TH12,TT13,DL14,DTW14}?
See \cite{TBR15} for recent progress on this last question.

\bigskip\textbf{Acknowledgements.} We acknowledge many discussions with our
colleagues regarding the strong converse for quantum capacity. This includes
Mario Berta, Frederic Dupuis, Will Matthews, Ciara Morgan, Naresh Sharma,
Stephanie Wehner, and Dong Yang. We are grateful to the anonymous
referees for many helpful comments about our paper. MT is funded by the Ministry of Education
(MOE) and National Research Foundation Singapore, as well as MOE Tier 3 Grant
\textquotedblleft Random numbers from quantum processes\textquotedblright%
\ (MOE2012-T3-1-009). MMW acknowledges startup funds from the Department of
Physics and Astronomy at LSU, support from the NSF through Award
No.~CCF-1350397, and support from the DARPA Quiness Program through US Army
Research Office award W31P4Q-12-1-0019. MMW\ is also grateful to the quantum
information theory group at the Universitat Aut\`{o}noma de Barcelona and to
Stephanie Wehner's quantum information group at the Centre for Quantum
Technologies, National University of Singapore, for hosting him for research
visits during May-June 2014. AW acknowledges financial support by the Spanish
MINECO, project FIS2008-01236 with the support of FEDER funds, the EC STREP
\textquotedblleft RAQUEL\textquotedblright, the ERC Advanced Grant
\textquotedblleft IRQUAT\textquotedblright, and the Philip Leverhulme Trust.

\appendices

\section{Proof of Lemma \ref{lm:continuity}}
\label{sec:ogawa-nagaoka}

\begin{IEEEproof}[Proof of Lemma \ref{lm:continuity}]
The first statement follows because the underlying sandwiched relative entropy, $\widetilde{D}_{\alpha}(\rho\|\sigma)$, is monotonically increasing in $\alpha$~\cite[Theorem 7]{MDSFT13} for all $\rho \in \mathcal{S}$ and $\sigma \in \mathcal{P}$, i.e.\
\begin{equation}
\widetilde{D}_{\alpha}(  \rho\Vert\sigma)  \geq\widetilde{D}%
_{\beta}(  \rho\Vert\sigma)
\end{equation}
for all $\alpha\geq\beta\geq 0$.
This already establishes that the limit exists and satisfies
\begin{equation}
  \lim_{\alpha \to 1^+} \widetilde{R}_{\alpha}(\mathcal{N}) \geq R(\mathcal{N}) .
\end{equation}

We would like to show the opposite inequality.
Consider the following bound~\cite[Lemma 8]{TCR09} 
(see also~\cite[Eq.~(21)]{WWY13})
\begin{equation}
\widetilde{D}_{1+\delta}(  \rho\Vert\sigma)  \leq D(  \rho
\Vert\sigma)  +4 \delta \left[  \log\nu\left(
\rho,\sigma\right)  \right]  ^{2},
\end{equation}
which holds for $\delta\in\left(  0,\log3/\left(  4\log\nu\left(
\rho,\sigma\right)  \right)  \right)  $ where%
\begin{equation}
\nu(  \rho,\sigma)  := \tr (\rho^{3/2} \sigma^{- 1/2})
+ \tr (\rho^{1/2} \sigma^{1/2}) +1.
\end{equation}
Let us for the moment assume that $\sigma > 0$ with $\tr(\sigma) \leq 1$, and its smallest eigenvalue is denoted as $\lambda$. In this case, we can bound
$ \nu(\rho,\sigma) \leq 2 + 1/{\sqrt{\lambda}}$.

 For any state $\rho_{RB}$ and $\delta > 0$ sufficiently small, 
we can apply the bound above to arrive at%
\begin{align}
&\inf_{\sigma_{RB}\in\RainsSet(  R:B)  }\widetilde{D}_{1+\delta}
( \rho_{RB} \Vert\sigma_{RB})  \\
&\qquad \leq \inf_{\tau_{RB}\in\RainsSet(  R:B)  }\widetilde{D}_{1+\delta}
(  \rho_{RB}  \Vert  (1- \delta) \tau
_{RB} + \delta  \pi_{RB} )  \\
&\qquad \leq \inf_{\tau_{RB}\in\RainsSet(  R:B)  } D(  \rho_{RB}  \Vert  (1-\delta) \tau
_{RB} + \delta \pi_{RB} ) \nonumber \\
& \qquad \qquad + 4 \delta \left( \log \left( 2 + \frac{\sqrt{|A| |B|}}{\sqrt{\delta}} \right) \right)^2 \\
& \qquad \leq%
\inf_{\tau_{RB}\in\RainsSet(  R:B)  } D(  \rho_{RB}  \Vert  \tau
_{RB} ) + \log \frac1{1-\delta} \nonumber \\
& \qquad \qquad + 4 \delta \left( \log \left( 2 + \frac{\sqrt{|A| |B|}}{\sqrt{\delta}} \right) \right)^2 \label{eq:uniform-bound}
\end{align}
The first inequality follows by picking $\sigma_{RB}$ to be of the form $\sigma_{RB} = \left(  1- \delta
\right)  \tau_{RB} + \delta \pi_{RB}$ where $\pi_{RB}$ is the fully mixed state on $RB$.
Also, note that $\left(  1-\delta\right)  \tau_{RB} + \delta \pi_{RB}\in \RainsSet(  R:B)$.
To verify the second inequality, note that the minimum eigenvalue of $\left(  1-\delta\right)  \tau_{RB} + \delta \pi_{RB}$ is always larger than $\delta/(|A| |B|)$. (The system $R$ can be chosen to be of size $|A|$ without loss of generality, where $|A|$ is the dimension of
the channel input.)  Finally, recall
that
$D(  \rho\Vert\sigma)  \leq D\left(  \rho
\Vert\sigma^{\prime}\right)$
whenever $\sigma^{\prime}\leq\sigma$ to verify the last inequality.

Since, crucially, the upper bound in~\eqref{eq:uniform-bound} is uniform in $\rho_{RB}$, we can immediately conclude that
\begin{multline}
  \widetilde{R}_{1+\delta}(\mathcal{N}) \leq R(\mathcal{N}) + \log \frac1{1-\delta} \\+ 4 \delta \left( \log \left( 2 + \frac{\sqrt{|A| |B|}}{\sqrt{\delta}} \right) \right)^2 \label{eq:nice-bound}
\end{multline}
by maximizing over channel output states as in the definition of $\widetilde{R}_{1+\delta}$ and $R$.
Thus, 
  $\lim_{\alpha \to 1^+} \widetilde{R}_{\alpha}(\mathcal{N}) \leq R(\mathcal{N})$,
concluding the proof.
\end{IEEEproof}

\section{Proof of Corollary \ref{cor:von-Neumann-subadd}}

\label{sec:appendix}

\begin{IEEEproof}
We can focus only on operators $\tau_{RB}$ such that supp$\left(  \rho
_{RB}\right)  \subseteq\ $supp$(  \tau_{RB})  $, where $\rho
_{RB}=\mathcal{N}_{A\rightarrow B}(  \phi_{RA})  $. This is because
the quantity of interest contains a minimization over all $\tau_{RB}$. For any sufficiently small $\delta>0$%
\begin{align}
R(  \mathcal{N}^{\otimes n})  \leq\widetilde{R}_{1+\delta
}(  \mathcal{N}^{\otimes n})  \leq\frac{1+\delta}{\delta}\left\vert A\right\vert ^{2}\log n+n\widetilde
{R}_{1+\delta}(  \mathcal{N}).
\end{align}
The first inequality is from the monotonicity of the sandwiched R\'{e}nyi
relative entropy in the R\'{e}nyi parameter \cite[Theorem 7]{MDSFT13}. The
second inequality follows from Theorem~\ref{thm:weak-subadditivity}.

This invites an application of~\eqref{eq:nice-bound}, which gives
\begin{multline}
  \frac{1}{n} R(  \mathcal{N}^{\otimes n}) \leq R(\mathcal{N}) + \frac{1+\delta}{\delta}\left\vert A\right\vert ^{2} \frac{\log n}{n} \\+ \log \frac1{1-\delta} + 4 \delta \left( \log \left( 2 + \frac{\sqrt{|A| |B|}}{\sqrt{\delta}} \right) \right)^2 
\end{multline}
Choosing $\delta = 1/\sqrt{n}$ and taking the limit $n \to \infty$ then immediately yields the desired result.
\end{IEEEproof}

\bibliographystyle{ieeetr}
\bibliography{Ref}

\begin{IEEEbiographynophoto}{Marco Tomamichel} (M'13) received the M.Sc.\ in Electrical Engineering and Information Technology degree from ETH Zurich (Switzerland) in 2007. He then graduated with a Ph.D.\ in Physics at the Institute of Theoretical Physics at ETH Zurich in 2012. From 2012 to 2014 he was a Research Fellow and then Senior Research Fellow at the Centre for Quantum Technologies at the National University of Singapore. Currently he is a Lecturer in the School of Physics at the University of Sydney.
His research interests include classical and quantum information theory with finite resources as well as applications to cryptography.
\end{IEEEbiographynophoto} 

\begin{IEEEbiographynophoto}{Mark M. Wilde} (M'99-SM'13) was born in Metairie, Louisiana, USA. He received the Ph.D. degree in electrical engineering from the University of Southern California, Los Angeles, California, in 2008. He is an Assistant Professor in the Department of Physics and Astronomy and the Center for Computation and Technology at Louisiana State University. His current research interests are in quantum Shannon theory, quantum optical communication, quantum computational complexity theory, and quantum error correction.
\end{IEEEbiographynophoto} 

\begin{IEEEbiographynophoto}{Andreas Winter}
received a Diploma degree in Mathematics from the Freie 
Universit\"at Berlin, Berlin, Germany, in 1997, and a Ph.D.
degree from the Fakult\"at f\"ur Mathematik, Universit\"at 
Bielefeld, Bielefeld, Germany, in 1999. 
He was Research Associate at the University of Bielefeld 
until 2001, and then with the Department of Computer Science 
at the University of Bristol, Bristol, UK. In 2003, still with 
the University of Bristol, he was appointed Lecturer in 
Mathematics, and in 2006 Professor of Physics of Information.  
Since 2012 he has been ICREA Research Professor with the Universitat
Aut\`{o}noma de Barcelona, Barcelona, Spain. 
\end{IEEEbiographynophoto}

\end{document}